\documentclass[12pt]{article}

\newcommand{\ra}{\rightarrow}
\newcommand{\nn}{\nonumber}

\newcommand{\norm}[1]{\|#1\|}

\newcommand{\Ima}{{\rm Im}}
\newcommand{\Rea}{{\rm Re}}
\newcommand{\bs}{\backslash}

\title{\vspace{-1in}\parbox{\linewidth}{\small\hfill\shortstack{IASSNS-HEP-98/26\\
CALT-68-2166}}
\vspace{0.6in}\\
Singular Monopoles and Gravitational Instantons}
\author{Sergey A.  Cherkis\thanks{Research supported in part by DOE grant 
DE-FG03-92-ER40701}\\
{\sl\small California Institute of Technology}\\
{\sl\small  Pasadena, CA 91125}
\and Anton Kapustin\thanks{Research supported in part by DOE grant DE-FG02-90-ER40542}\\
{\sl\small School of Natural Sciences, Institute for Advanced Study}\\
{\sl\small Olden Lane, Princeton, NJ 08540}}

\begin{document}
\begin{titlepage}
\renewcommand{\thepage}{ }
\renewcommand{\today}{ }
\thispagestyle{empty}

\maketitle

\begin{abstract}
We model $A_k$ and $D_k$ asymptotically locally flat gravitational instantons
on the moduli spaces of solutions of $U(2)$ Bogomolny equations with
prescribed singularities. We study
these moduli spaces using Ward correspondence and find their twistor description.
This enables us to write down the K\"ahler potential for $A_k$ and $D_k$ gravitational
instantons in a relatively explicit form.

\end{abstract}
\end{titlepage}

\section{Introduction}
A gravitational instanton is a smooth four-dimensional manifold with a
Riemannian metric satisfying Einstein equations. A particularly interesting
class of gravitational instantons is that of four-dimensional 
hyperk\"ahler manifolds, i.e. manifolds with holonomy group contained in
$SU(2)$. 
A hyperk\"ahler manifold can be alternatively
characterized as a Riemannian manifold admitting three covariantly
constant complex structures $I,J,K$ satisfying the quaternion relations
\begin{equation}
IJ=-JI=K, {\rm etc.}
\end{equation}
such that the metric is Hermitian with respect to $I,J,K$. 
Covariant constancy of $I,J,K$ implies that three 2-forms 
$\omega_1=g(I\cdot,\cdot), \omega_2=g(J\cdot,\cdot), \omega_3=g(K\cdot,\cdot)$ are 
closed.
If we pick one of the complex structures, say $I$, we may regard a 
hyperk\"ahler manifold as a complex manifold equipped with K\"ahler
metric (with K\"ahler form $\omega_1$) and a complex symplectic
form $\omega=\omega_2+i\omega_3$.

Hyperk\"ahler four-manifolds arise in several physical problems. For example, 
compactification of string and M-theory on hyperk\"ahler four-manifolds
preserves one half of supersymmetries and provides exact solutions of 
stringy equations of motion. 

The only compact hyperk\"ahler four-manifolds are $T^4$ and K3, but the 
K3 metric is not known explicitly. In the noncompact case
there are several possibilities to consider. The are no nontrivial
hyperk\"ahler metrics asymptotically approaching that of ${\bf R}^4$, but the
situation becomes more interesting if one asks that the metric be
only Asymptotically Locally Euclidean (ALE), i.e. that the metric look asymptotically
like the quotient of ${\bf R}^4$ by a finite group of isometries.
All such metrics fit into the ADE classification of Kronheimer which we
now briefly explain. Let $\Gamma$ be a finite subgroup of $SU(2)$.
There is a natural correspondence between such $\Gamma$'s and 
ADE Dynkin diagrams: $A_k$ diagram corresponds to the cyclic group
${\bf Z}_{k+1}$, $D_k$ diagram corresponds to the binary dihedral group ${\bf D}_{k-2}$ of order 
$4(k-2)$, and $E_k$ diagrams correspond to symmetry groups of
tetrahedron, cube, and icosahedron. Since $SU(2)$ acts on ${\bf C}^2$ by
the fundamental representation, we may consider quotients ${\bf C}^2/\Gamma$
(known as Kleinian singularities). Kronheimer showed that resolutions
of Kleinian singularities admit ALE hyperk\"ahler metrics, and that
all such metrics arise in this way~\cite{ALE,Torelli}. In the $A_k$ case the metric
has been known explicitly for some time: it is the Gibbons-Hawking
metric with $k+1$ centers~\cite{GH}. Kronheimer provided an implicit 
construction of $D_k$ and $E_k$ ALE gravitational instantons as
hyperk\"ahler quotients~\cite{ALE}.

Another interesting class of noncompact gravitational instantons
is that of Asymptotically Locally Flat (ALF) manifolds. This means
that the metric asymptotically approaches the metric on $({\bf S}^1\times {\bf R}^3)/\Gamma$
where $\Gamma$ is some finite group. The only known hyperk\"ahler
metric of this sort is the multi-Taub-NUT metric. As a complex manifold
the $k+1$-center multi-Taub-NUT space is isomorphic to the resolution of 
${\bf C}^2/{\bf Z}_{k+1}$,
so we will call it $A_k$-type ALF gravitational instanton. Compactification
of M-theory of this manifold is equivalent to a configuration of
$k+1$ parallel D6 branes in IIA string theory. Furthermore,
it is expected that a configuration of an O6+ orientifold and $k$ D6 branes
in IIA string theory corresponds to the compactification of M-theory on
a $D_k$ ALF space~\cite{Sen}, i.e. an ALF gravitational instanton isomorphic to the resolution 
of ${\bf C}^2/{\bf D}_{k-2}$. More generally, any compactification of
M-theory on an ALF hyperk\"ahler manifold should correspond to a IIA brane
configuration preserving half of supersymmetries. Thus it is of interest to find
all four-dimensional ALF hyperk\"ahler metrics in as explicit form as
possible.

In our previous paper~\cite{us3} we constructed $D_k$ ALF metrics from moduli
spaces of certain ordinary differential equations (Nahm equations).
In this paper we construct both $A_k$ and $D_k$ ALF hyperk\"ahler four-mani\-folds
from moduli spaces of solutions of $U(2)$ Bogomolny equations on ${\bf R}^3$ with
prescribed singularities. Solutions of $SU(2)$ Bogomolny equations with singularities
were previously considered by Kronheimer~\cite{KrTh}, and much of our discussion closely
follows that in Ref.~\cite{KrTh}. 
The idea is that, on one hand, the moduli space of Bogomolny equations
carries natural hyperk\"ahler structure, while on the other hand the
solutions can be found by means of Ward correspondence.
This approach yields directly the twistor space (in the sense of Penrose)
of the moduli space of solutions. To get the metric itself one needs 
to find an appropriate family of sections of the twistor space. In both $A_k$
and $D_k$ cases we were able to identify the correct family of sections
only modulo some finite choices. The $A_k$ metrics are simple enough so that 
one can explicitly see that only one choice gives nonsingular metrics. 
In the end of subsection 5.1 we argue that the $D_k$ twistor spaces we have constructed
correspond to everywhere smooth hyperk\"ahler metrics as well. We also show
that the $D_k$ ALF metrics we obtain are identical to those found in Ref.~\cite{us3}. 

Let us explain what we mean by solutions of Bogomolny equations with prescribed
singularities. Recall that
Bogomolny equations on ${\bf R}^3$ are equations for a connection $A$ in a vector bundle
$B$ over ${\bf R}^3$ and a section $\Phi$ of ${\rm End} B$:
$$*F_A=D_A\Phi.$$
Let $\norm{a}$ be the Ad-invariant norm on $u(2),$ $\norm{a}^2=-\frac{1}{2}
{\rm Tr}{a^2}$. Fix $k$ distinct points ${\vec p}_1,\ldots,{\vec p}_k\in {\bf R}^3$.
\emph{A singular $U(2)$ monopole} is a solution of $U(2)$ Bogomolny
equations on ${\bf R}^3\bs\{\vec{p}_1,\ldots,\vec{p}_k\}$ satisfying the following conditions.

(i) As $\vec{r}\ra\vec{p}_\alpha$ $\,2r_\alpha\Phi\ra i\ {\rm diag}\left(0,-\ell_\alpha\right)$
up to gauge transformations, and
$d\left(r_\alpha\norm{\Phi}\right)$ is bounded. Here $r_\alpha=|\vec{r}-\vec{p}_\alpha|$, 
$\alpha=1,\ldots,k$.

(ii) As $r\ra\infty$ one has asymptotic expansions, up to gauge transformations,
$$\Phi=i\ {\rm diag}\left(\mu_1-\frac{n}{2r},\mu_2+\frac{n-\sum \ell_\alpha}{2r}\right)
+O(1/r^2),$$
$$\frac{\partial\norm{\Phi}}{\partial\Omega}=O\left(1/r^2\right),\ \norm{D\Phi}=
O\left(1/r^2\right).$$
We will refer to $n$ as the nonabelian charge of the monopole, and to $\{\ell_\alpha\}$
as its abelian charges. We will assume that $\mu_1>\mu_2$. We also set $n'=n-\sum_\alpha
\ell_\alpha,\mu=\mu_1-\mu_2,$ for short. 

Every fiber of the complex rank two bundle $B$ splits into the eigenspaces of $\Phi$,
$B=M_1\oplus M_2$, near $\vec{r}=\vec{p}_\alpha$ or when $r\ra\infty$.
Let $M_1$ correspond the eigenvalue of
$\Phi$ diverging as $\vec{r}\ra\vec{p}_\alpha.$
It is a simple consequence of Bogomolny equations that $-\ell_\alpha$ is the degree of $M_1$
restricted to a small 2-sphere around $\vec{r}=\vec{p}_\alpha$. Similarly, $-n$ and $n'$
are the degrees of eigensubbundles of $B$ restricted to a large 2-sphere. Therefore $n$
and $\{\ell_\alpha\}$ are integers.

String theory considerations imply that the moduli space of $n=1$ mo\-no\-pole with 
$\ell_\alpha=1,\alpha=1,\ldots,k$ is an $A_{k-1}$ ALF gravitational instanton, and the 
centered moduli space of $n=2$ monopole\footnote{The centered $n=2$ monopole moduli space is 
a $U(1)$ hyperk\"ahler quotient of the $n=2$ monopole moduli space; see section 5 below.}
with $\ell_\alpha=1,\alpha=1,\ldots,k$ is a $D_k$ ALF gravitational instanton~\cite{us}.
In this paper we show that this is indeed the case. The main tool is the Ward correspondence
described in section 2. In section 3 we use it to derive the twistor space for
arbitrary $n$. In sections 4 and 5 we deal with $n=1$ and $n=2$ cases, respectively.
We show that the moduli spaces are resolutions of $A_{k-1}$ and $D_k$ singularities,
as expected, find the real holomorphic sections of the twistor spaces, and derive 
the K\"ahler potentials for the metrics using the generalized Legendre transform
method of Refs.~\cite{LR,IR}.

\section{Ward correspondence}

{}From now on we restrict ourselves to the case $\ell_\alpha=1,\alpha=1,\ldots,k$.
We make some comments on the more general case
of positive $\ell_\alpha$ at the end of subsection 5.1.
To construct the moduli space of singular $U(2)$ monopoles, we will use
a version of Ward correspondence due to Hitchin~\cite{H1}. The set of
all oriented straight lines ${\bf T}$ in ${\bf R}^3$ has a natural complex structure, as it is
the tangent bundle of the projective line. 
${\bf T}$ can be covered by two patches $V_0(\zeta\neq\infty)$ and
$V_1(\zeta\neq 0)$ with coordinates $(\eta,\zeta)$ and $(\eta',\zeta')=(\eta/\zeta^2,1/\zeta)$.
For any point $\vec{x}\in {\bf R}^3$ the set of all oriented straight lines through $\vec{x}$ 
sweeps out a projective line $P_x\in {\bf T}$; thus there is a holomorphic map 
$P_x:{\bf P}^1\ra {\bf T}$. The reversal of the orientation of lines in ${\bf R}^3$ is an 
antiholomorphic map $\tau:{\bf T}\ra {\bf T}$ 
satisfying $\tau^2=id$. It is called the real structure of ${\bf T}$. For any $\vec{x}$
it acts on $P_x$ as the antipodal map. Thus $P_x$ is a real holomorphic section of ${\bf T}$.

For any straight line in ${\bf R}^3$ 
$$\gamma=\left\{\vec{x}|\vec{x}=\vec{u}t+\vec{v},\vec{u}\cdot\vec{u}=1,
\vec{u}\cdot\vec{v}=0\right\}$$ let
\begin{eqnarray}
\gamma_+&=&\left\{\vec{x}|\vec{x}=\vec{u}t+\vec{v},t>R\right\},\\ \nn
\gamma_-&=&\left\{\vec{x}|\vec{x}=\vec{u}t+\vec{v},t<-R\right\},
\end{eqnarray}
where $R$ is a positive number greater than any $|\vec{p}_\alpha|$. 
Now we define two complex rank 2 vector bundles $E^+$ and $E^-$ over ${\bf T}$:
\begin{eqnarray}
E^+&=&\left\{s\in \Gamma\left(\gamma_+,E\right)|D_\gamma s=i\Phi s\right\}, \\ \nn
E^-&=&\left\{s\in \Gamma\left(\gamma_-,E\right)|D_\gamma s=i\Phi s\right\}.
\end{eqnarray} 
{}From Bogomolny equations it follows, as in Ref.~\cite{H1}, that these bundles are holomorphic.
The real structure $\tau$ on ${\bf T}$ can be lifted to an antilinear antiholomorphic map
$$\sigma: E^+\ra (E^-)^*.$$ Thus every solution of $U(2)$ Bogomolny equations maps to a pair
of holomorphic rank two bundles on ${\bf T}$ interchanged by the real structure.

Let $P_x$ denote the real section corresponding to $\vec{x}$, and $P_\alpha$ the real section
corresponding to $\vec{p}_\alpha$. Let $P$ be the union of all $P_\alpha$. If $\gamma$ does
not pass through any of $\vec{p}_\alpha$, any solution $s$ can be continued from
$\gamma_+$ to $\gamma_-$. This defines a natural indentification of the fibers
$E^+_\gamma$ and $E^-_\gamma$. Therefore we have an isomorphism
\begin{equation}\label{h}
h:E^+|_{{\bf T}\bs P}\ra E^-|_{{\bf T}\bs P}.
\end{equation}

For nonsingular monopoles $h$ extends to an isomorphism over the whole ${\bf T}$, therefore
the Ward correspondence maps a nonsingular monopole to a holomorphic bundle over ${\bf T}$.
In the present case $h$ or $h^{-1}$ may have singularities at $P$, and the Ward correspondence maps
a singular monopole into a triplet $(E^+,E^-,h)$. This triplet satisfies a certain
triviality constraint which we now proceed to formulate.

For any $\vec{x}$ distinct from all $\vec{p}_\alpha$ the intersection $P_x\bigcap P$
consists of an even number of points. For a generic
$\vec{x}$  the cardinality of $Q_x=P_x\bigcap P$ is $2k$. For any $\vec{x}$ we can 
arbitrarily split $Q_x$ into two sets of equal cardinality
$Q_x^+$ and $Q_x^-$ and construct a vector bundle $E_x$ over
$P_x$ by gluing together $E^+$ restricted to $P_x\bs Q_x^+$ and $E^-$ restricted to 
$P_x\bs Q_x^-$, with the transition function $h$. (Of course, $E_x$ depends on the splitting.)
The triviality constraint is that for any $\vec{x}$ there is a splitting 
$Q_x=Q_x^+\bigcup Q_x^-$ such that $E_x$ is trivial.

Now we state the Ward correspondence between singular $U(2)$ monopoles and twistor data.
There is a bijection\footnote{The injectivity of Ward correspondence can be shown by a
straightforward modification of the argument in Ref.~\cite{H1}. We conjecture the surjectivity
by analogy with the nonsingular case.}
between singular monopoles modulo gauge transformations and pairs
$(E^+,E^-)$ of holomorphic rank 2 bundles over ${\bf T}$ equipped with an isomorphism
Eq.~(\ref{h}) satisfying the following conditions:

(a) For any $\vec{x}\neq \vec{p}_\alpha$ there is a splitting $Q_x=Q_x^+\bigcup Q_x^-$
such that $E_x$ is trivial.

(b) In the vicinity of each point of $P$ there exist trivializations of 
$E^+$ and $E^-$ such that $h$ takes the form
\begin{equation}\label{eta}
h=\left(\begin{array}{ccc}
1 & 0 \\ 0 & \prod_\alpha(\eta-P_\alpha(\zeta)) \\ 
\end{array}\right),
\end{equation}
so that $h$ extends across $P$ to a morphism $E^+\ra E^-$.

(c) The real structure $\tau$ on ${\bf T}$ lifts to an antilinear antiholomorphic map
$\sigma: E^+\ra (E^-)^*$.

Let us explain where (a) and (b) come from.
The condition (b) arises from studying the behavior of the solutions of 
the equation $D_\gamma s=i\Phi s$ as $\gamma$ approaches $\vec{p}_\alpha$. 
Details can be found in Ref.~\cite{KrTh}. (There the $SU(2)$ case was analyzed, but 
the extension to $U(2)$ is straightforward).
To demonstrate (a) it is sufficient to exhibit a holomorphic trivialization of $E_x$.
Take any ${\vec x}\neq {\vec p}_\alpha,\alpha=1,\ldots,k$ and recall that $P_x$ consists
of all straight lines 
$\gamma$ passing through $\vec{x}$. To obtain a holomorphic section of $E_x$ pick
a vector $v_1$ in the fiber of $B$ over ${\vec x}$ and take it as an initial condition
for the equation $D_\gamma s=i\Phi s$ at $t=0$. Integrating it forward and backward in 
$t$ and varying $\gamma$
yields sections of $E^+$ and $E^-$ related by $h$. It is easy to check that they are 
holomorphic and thereby combine into a holomorphic section $s_1$ of $E_x$. 
To get a section $s_2$ of $E_x$ linearly independent from $s_1$ just pick a vector $v_2$ 
linearly independent from $v_1$ and repeat the procedure. (This argument has to be 
modified if there is a straight line $\gamma$ passing through $\vec{x}$ and 
$\alpha,\beta\in \{1,\ldots,k\}$ such that $\vec{p}_\alpha$ and $\vec{p}_\beta$ lie
on $\gamma$ and $\vec{x}$ separates them. In this case one of the vectors $v_1,v_2$ 
has to be varied, $v_i\sim \zeta^{-1}$, as one varies $\gamma$.) 

We now want to encode the twistor data in an algebraic curve $S\subset {\bf T}$, 
in the spirit of Ref.~\cite{H1}. We denote by ${\cal O}(m)$ the pullback to ${\bf T}$ of
the unique degree $m$ line bundle on ${\bf P}^1$, and by $L^x(m)$ a line bundle over 
${\bf T}$ with the transition function $\zeta^{-m} e^{-x\eta/\zeta}$ from $V_0$ to $V_1$.
Let $L_1^+$ be a line subbundle of $E^+$ which consists of
solutions of $D_\gamma s=i\Phi s$ bounded by $const\cdot\exp(-\mu_1t) t^n$ as $t\ra +\infty$.
Similarly, a line bundle $L_1^-\subset E^-$ consists of solutions bounded by 
$const\cdot\exp(-\mu_2t) t^{-n'}$ as $t\ra -\infty$. 
The line bundles $L_2^+$ and $L_2^-$ are defined by $$L_2^+=E^+/L_1^+,\qquad L_2^-=E^-/L_1^-.$$
As in Ref.~\cite{H1} the asymptotic conditions on the Higgs field can be used to show that
$L_{1,2}^+$ and $L_{1,2}^-$ are holomorphic line bundles, and that the following isomorphisms 
hold: $$L_1^+\simeq L^{\mu_1}(-n),\ L_2^+\simeq L^{\mu_2}(n'),\ L_1^-\simeq L^{\mu_2}(-n'),
\ L_2^-\simeq L^{\mu_1}(n).$$

Consider a composite map
$$\psi: L_1^+\ra E^+\ra E^-\ra L_2^-,$$
where the first arrow is an inclusion, the second arrow is $h$, and the third arrow is a 
natural projection.
We may regard $\psi$ as an element of  $H^0({\bf T},{\cal O}(2n))$.
Let us define the spectral curve $S$ to be the zero level of $\psi$. $S$ is in the linear 
system ${\cal O}(2n)$. Arguments identical to those in Ref.~\cite{H1}
can be used to prove that $S$ is compact and real (i.e., $\tau(S)=S$). 

Consider now a map $\phi:\wedge^2 E^+\ra \wedge^2 E^-$ induced by $h$. By virtue of Eq.~(\ref{eta})
the zero level of $\phi$ is precisely $P$. We will assume in what 
follows that $S$ does not contain any of $P_\alpha$
as components. Physically this corresponds to the requirement that none of the 
nonabelian monopoles is located at $\vec{x}=\vec{p}_\alpha$. For simplicity we will
also assume that $S\bigcap P$ consists of $2nk$ points (this is a generic situation).

The construction here bears a close resemblance to that in Ref.~\cite{HM}, where 
nonsingular monopoles for all classical groups were constructed. According to 
Ref.~\cite{HM}, the spectral
data for a nonsingular $SU(3)$ monopole with magnetic charge $(k,n)$ include a pair 
of spectral curves $S_1,S_2$ in the linear systems ${\cal O}(2n),{\cal O}(2k)$. 
Our $S$ and $P$ are analogs of $S_1$ and $S_2$. 
The condition that $S\bigcap P$ consists of $2nk$ points is analogous
to the requirement in Ref.~\cite{HM} that the monopoles are generic.
(This resemblance is not a coincidence: if we consider an $SU(3)$ gauge theory 
broken down to $SU(2)\times U(1)$ by a large vev of an adjoint Higgs field, the $(k,n)$ 
monopoles of $SU(3)$ reduce to singular
monopoles of $SU(2)\times U(1)$ with nonabelian charge $n$ and total abelian charge $k$.
In this limit the spectral data of Ref.~\cite{HM} must reduce to ours.)

Since $L_1^+|_S={\rm ker}\ \psi|_S,$ we have a well-defined holomorphic map
$\rho: L_2^+|_S\ra L_2^-|_S$ induced by $h$. There is also a 
holomorphic map $\xi: L_1^+|_S\ra L_1^-|_S$ induced by $h$. Thus we have natural elements 
$\rho\in H^0\left(S,L^{\mu}(k)\right)$ and $\xi\in H^0\left(S,L^{-\mu}(k)\right).$
It also easily follows from the definition that $\rho\otimes\xi=\phi|_S$, and therefore
the divisors of both $\rho$ and $\xi$ are subsets of $S\bigcap P$. $\rho$ and
$\xi$ are interchanged by real structure, and therefore the same is true about their
divisors. It follows that the divisors of $\rho$ and $\xi$ are disjoint and have equal
cardinality. Thus we can define the spectral data for a generic singular monopole to consist of

(i) A spectral curve $S$, which is a real compact curve in the linear system ${\cal O}(2n)$ such
that $S\bigcap P$ consists of $2nk$ disjoint points.

(ii) A splitting $S\bigcap P=Q^+\bigcup Q^-$ into sets of equal cardinality interchanged by
$\tau$.

(iii) A section $\rho$ of $L^{\mu}(k)|_S$ with divisor $Q^+$ and a section
$\xi$ of $L^{-\mu}(k)|_S$ with divisor $Q^-$. $\rho$ and $\xi$ are interchanged by real
structure. 

The condition (iii) is a constraint on $S$. It implies that $\rho$ and $\xi$ satisfy
\begin{equation}\label{master}
\rho\xi=\prod_\alpha (\eta-P_\alpha(\zeta)).
\end{equation}
For nonsingular monopoles it reduces
to the requirement that $L^{\mu}|_S$ be trivial, as in Ref.~\cite{H1}.
As a consequence of (iii), $L^{2\mu}|_S\left[Q_--Q_+\right]$ is trivial.

Recall that the spectral data for nonsingular $SU(2)$ monopoles satisfy an additional 
constraint, the ``vanishing theorem'' of Ref.~\cite{H2}.
It says that  $L^{z\mu}(n-2)$ is nontrivial for $z\in (0,1).$ A natural guess for the
analogue of this condition in our case is 

(iv) $L^{z\mu}(n-2)\left[-Q_+\right]$ is nontrivial for $z\in (0,1).$

We already mentioned a close connection of the spectral data for singular $U(2)$ monopoles
and those for nonsingular $SU(3)$ monopoles~\cite{HM} with the largest Higgs vev set to 
$+\infty.$ Consequently, one can obtain the condition (iv) from the ``vanishing theorem''
of Ref.~\cite{HM} by taking the appropriate limit. A direct derivation of (iv) should 
also be possible.

Arguments very similar to those in Ref.~\cite{H1} show that the spectral data determine 
the singular monopole uniquely. A natural question
is if there is a one-to-one correspondence between singular $U(2)$ monopoles and
spectral data defined by (i-iv). The answer was positive for nonsingular $SU(2)$ 
monopoles~\cite{H2}, so it is highly plausible that the same is true in the present case.
Presumably a proper proof of this can be achieved by converting the spectral data into
solutions of Nahm equations, and then reconstructing singular monopoles by an inverse Nahm
transform~\cite{H2,HM}.

\section{Twistor space for singular monopoles}

Having established the correspondence between singular $U(2)$ monopoles and algebraic data on 
${\bf T}$, we now proceed to construct the twistor space $Z_n$ for the
moduli space of a singular monopole with nonabelian charge $n$. We follow 
the method of Ref.~\cite{AH}. For fixed $\zeta=\zeta_0$ every point in $Z_n$ yields a 
spectral curve $S$ which intersects the fiber of ${\bf T}$ over $\zeta_0$ at $n$ points. 
Thus we have a projection $$Z_n\ra \oplus_{j=1}^n {\cal O}(2j)=Y_n.$$
Concretely, if $S$ is given by $\eta^n+\eta_1\eta^{n-1}+\cdots+\eta_n=0$, the corresponding 
point in $Y_n$ is $(\eta_1,\ldots,\eta_n)$. Now consider an $n$-fold cover of $Y_n$
$$X_n=\left\{(\eta,\eta_1,\ldots,\eta_n)\in {\cal O}(2)\oplus Y_n|\eta^n+\eta_1\eta^{n-1}+
\cdots+\eta_n=0\right\}.$$
There are two natural projections $\pi_1:X_n\ra {\bf T}$ and $\pi_2:X_n\ra Y_n$. Using these 
projections, we get a rank $n$ bundle $V^+$ over $Y_n$ as a direct image sheaf 
$V^+=\pi_{2*}\pi_1^*L^{\mu}(k)$. Similarly, we get a rank $n$ bundle 
$V^-=\pi_{2*}\pi_1^*L^{-\mu}(k)$. For any point in $Z_n$ we have a section $\rho$ of 
$L^{\mu}(k)|_S$ and a section $\xi$ of $L^{-\mu}(k)|_S$. Therefore,
there is an inclusion $Z_n\subset V^+\oplus V^-$. To describe this inclusion more concretely,
we must rewrite the condition (iii) in terms of sections of $V^\pm$.
The result is as follows. Let $U$ be a $2n+1$-dimensional subvariety in ${\bf C}^{3n+1}$ with coordinates 
$(\zeta,\eta_1,\ldots,\eta_n,\rho_0,\ldots,\rho_{n-1},\xi_0,\ldots,\xi_{n-1})$ defined by 
\begin{eqnarray}\label{direct}
&&(\rho_0+\rho_1\eta+\cdots+\rho_{n-1}\eta^{n-1})(\xi_0+\xi_1\eta+\cdots+\xi_{n-1}\eta^{n-1})=
\prod_\alpha (\eta-P_\alpha (\zeta)), \nn \\
&&{\rm mod}\ \eta^n+\eta_1\eta^{n-1}+\cdots+\eta_n=0.
\end{eqnarray}
Take two copies of $U$ and glue them together over $\zeta\neq 0,\infty$ by
\begin{eqnarray}\label{trk}
&&\hspace{-15pt}\tilde\zeta=\zeta^{-1},\\ \nn
&&\hspace{-15pt}\tilde\eta_j=\zeta^{-2j}\eta_j,\ j=1,\ldots,n,\\ \nn
&&\hspace{-15pt}\tilde\rho_0+\tilde\rho_1\tilde\eta+\cdots+\tilde\rho_{n-1}\tilde\eta^{n-1}=
e^{-\mu\eta/\zeta}\zeta^{-k}(\rho_0+\rho_1\eta+\cdots+\rho_{n-1}\eta^{n-1}),\\ \nn
&&\hspace{-15pt}\tilde\xi_0+\tilde\xi_1\tilde\eta+\cdots+\tilde\xi_{n-1}\tilde\eta^{n-1}=
e^{\mu\eta/\zeta}\zeta^{-k}(\xi_0+\xi_1\eta+\cdots+\xi_{n-1}\eta^{n-1}),
\end{eqnarray}
all modulo $\eta^n+\eta_1\eta^{n-1}+\cdots+\eta_n=0$.
The resulting $2n+1$-dimensional variety is $Z_n$, the twistor space of singular monopoles with
nonabelian charge $n$.

To reconstruct the hyperk\"ahler metric from the twistor space one has to find a holomorphic 
section of $\Lambda^2T_F^*\otimes {\cal O}(2),$ where $T_F^*$ is the cotangent bundle of the fiber 
of $Z_n.$ Upon restriction to
any fiber of $Z_n$ this section must be closed and nondegenerate. 
An obvious choice (the same as in Ref.~\cite{AH}) is
\begin{equation}
\omega=4\sum_{j=1}^n \frac{d\rho(\beta_j)\wedge d\beta_j}{\rho(\beta_j)},
\end{equation}
where $\beta_j,j=1,\ldots,n$ are the roots of $\eta^n+\eta_1\eta^{n-1}+\cdots+\eta_n=0$.

\section{Moduli space $M_1$ of $n=1$ monopole}\label{k1}
Specializing the formulas of the previous section to $n=1$, we get that the twistor space
$Z_1$ is a hypersurface in the total space of $L^\mu(k)\oplus L^{-\mu}(k)$
\begin{equation}\label{eqk1}
\rho_0\xi_0=\prod_{\alpha=1}^k (\eta-P_\alpha(\zeta)),
\end{equation}
where $\rho_0\in L^\mu(k), \xi_0\in L^{-\mu}(k)$, and $\eta\in {\cal O}(2)$. 
Obviously, for fixed $\zeta$ this is a resolution of ${\bf C}^2/{\bf Z}_k$, so the corresponding
hyperk\"ahler metric is an $A_{k-1}$ gravitational instanton. In fact, it is well known
what the metric is: it is the multi-Taub-NUT metric with $k$ centers. In the remainder
of this section we rederive this result using the Legendre transform method of 
Refs.~\cite{HKLR,LR,IR}. This will serve as a warm-up for the 
discussion of $D_k$ ALF metrics in the next section.

First we find the real holomorphic sections of the twistor space $Z_1$. 
This amounts to solving Eq.~(\ref{eqk1}) with $\rho_0,\xi_0$, and $\eta$ now 
regarded as holomorphic sections
of the appropriate bundles. Recalling that $\eta=a\zeta^2+2b\zeta-\bar{a}$ and 
$P_{\alpha}(\zeta)=a_{\alpha}\zeta^2+
2 b_{\alpha}\zeta -\bar{a}_{\alpha}$ with $b,b_\alpha\in {\bf R}$, one gets in 
the patch $V_0$
\begin{eqnarray}
\rho_0&=&A e^{\textstyle +\mu(b+a\zeta)}\prod_{\alpha=1}^k \left(\zeta-u_{\alpha}\right),\nn \\
\xi_0&=&B e^{\textstyle -\mu(b+a\zeta)}\prod_{\alpha=1}^k \left(\zeta-v_{\alpha}\right),\nn
\end{eqnarray}
with $A B=\prod \left(a-a_{\alpha}\right)$. Here $u_\alpha$ and $v_\alpha$ are the roots 
of the equation $\eta(\zeta)=P_\alpha(\zeta),$
\begin{eqnarray}\label{uv}
u_{\alpha}&=&\frac{-(b-b_{\alpha})-\Delta_{\alpha}}{a-a_{\alpha}}, \nn\\
v_{\alpha}&=&\frac{-(b-b_{\alpha})+\Delta_{\alpha}}{a-a_{\alpha}},
\end{eqnarray}
with $\Delta_\alpha=\sqrt{(b-b_\alpha)^2+|a-a_\alpha|^2}>0.$ (The ambiguity in the sign of
$\Delta_\alpha$ is fixed by requiring that the hyperk\"ahler metric on this family of
sections be everywhere nonsingular. This is equivalent to asking that the normal bundle
of every section in the family be ${\cal O}(1)\oplus {\cal O}(1)$.)  
Since the real structure must interchange $\rho_0$ and $\xi_0$, we get
\begin{equation}
\label{real}
B\overline{B}=\prod \left(b-b_{\alpha}+\Delta_{\alpha}\right).
\end{equation}
Thus we have a family of solutions to Eq.~(\ref{eqk1}) parametrized by
$\Rea\ a,\ \Ima\ a,\ b,$ and ${\rm Arg}\ B$. 

Having found the real holomorphic sections, we compute the K\"ahler potential. 
The twistor space $Z_1$ is
fibered over ${\bf P}^1$ with an intermediate projection
$$Z_1\ra {\cal O}(2)\ra {\bf P}^1.$$
In the above $\zeta$ and $\eta$ are coordinates on the base and the fiber of ${\cal O}(2)$, 
respectively. The holomorphic 2-form $\omega\in \Lambda^2T^*\otimes {\cal O}(2)$
is given by
\begin{equation}
\omega=4 d\eta\wedge \frac{d\rho}{\rho}.
\end{equation}
For $\zeta\neq\infty$ we can choose $\eta(\zeta)$ and $\chi=2\log \frac{\rho}{\xi}$ as 
two coordinates on the moduli space $M_1$ holomorphic with respect to the complex
structure defined by $\zeta$. The coordinates in the patch $\zeta\neq 0$ are related 
to these as
\begin{equation}
\eta'=\eta/\zeta^2, \chi'=\chi-4\mu\eta/\zeta.
\end{equation}
The second equation here follows from $\rho_0$ and $\xi_0$ being sections of 
$L^{\mu}(k)$ and $L^{-\mu}(k)$. In terms of these coordinates
\begin{equation}
\omega=d\eta\wedge d\chi=\zeta^2 d\eta'\wedge d\chi'.
\end{equation}

Following Ref.~\cite{IR} we define an auxiliary function $\hat{f}$ and a contour $C$ by the equation
\begin{equation}
\oint_C \frac{d\zeta}{\zeta^j}\hat{f}=\oint_0\frac{d\zeta}{\zeta^j}\chi
+\oint_\infty\frac{d\zeta}{\zeta^j}\chi'=
\left(\oint_0+\oint_{\infty}\right)\frac{d\zeta}{\zeta^j}\chi-
4\mu\oint_{\infty}\frac{d\zeta}{\zeta^{j+1}}\eta.
\end{equation}
for any integer $j$. Here and in what follows the integrals $\oint_0$ and $\oint_\infty$ 
are taken along small positively oriented contours around respective points. This 
implies that in the first of this integrals the contour runs counterclockwise, while 
in the second one it runs clockwise. Substituting an explicit expression for $\chi$ we find
\begin{equation}\label{fhat1}
\oint_C \frac{d\zeta}{\zeta^j}\hat{f}=\sum_\alpha\oint_{\Gamma_\alpha} \frac{d\zeta}{\zeta^j}\ 
2\log\left(\eta(\zeta)-P_{\alpha}(\zeta)\right)+4\mu\oint_0 \frac{d\zeta}{\zeta^{j+1}}\eta.
\end{equation}
Here $\Gamma_\alpha$ is a figure-eight-shaped contour enclosing $u_\alpha$ and $v_\alpha$ (see
Figure 1).
\begin{figure}\label{fig1}
\setlength{\unitlength}{0.6em}
\begin{center}

\begin{picture}(17,8)

\qbezier(0,5)(-3.464,3)(0,1)
\qbezier(0,5)(0.884,5.5104)(7,3)
\qbezier(0,1)(0.884,0.4896)(7,3)
\qbezier(14,5)(13.116,5.5104)(7,3)
\qbezier(14,1)(13.116,0.4896)(7,3)
\qbezier(14,5)(17.464,3)(14,1)
\put(5.9,3.5){\line(-3,2){1.5}} \put(5.9,3.5){\line(-1,0){1.8}}
\put(1,3){\circle*{0.2}} \put(13,3){\circle*{0.2}} 
\put(1.6,3){\makebox(0,0)[l]{$u_\alpha$}}
\put(13,3){\makebox(-0.5,0)[r]{$v_\alpha$}}
\put(13,6){\makebox(0,0)[b]{$\Gamma_\alpha$}}

\end{picture}\end{center}
\caption{The contour $\Gamma_\alpha$ enclosing $u_\alpha$ and $v_\alpha$.}
\end{figure}
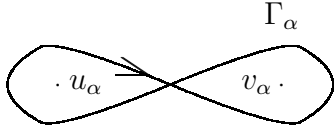
We define a function $G(\eta,\zeta)$ by $\partial G/\partial \eta=\hat{f}$. According to
Ref.~\cite{IR} the Legendre transform of the K\"ahler potential is given by
\begin{equation}
F(a,b)=\frac{1}{2\pi i}\oint_C \frac{d\zeta}{\zeta^2} G(\eta,\zeta).
\end{equation}
Using Eq.~(\ref{fhat1}) we find
\begin{equation}\label{F1}
F(a,b)=\frac{2\mu}{2\pi i}\oint_0 \frac{d\zeta}{\zeta^3}\eta^2+\sum_{\alpha=1}^k 
\frac{1}{2\pi i}\oint_{\Gamma_{\alpha}}\frac{d\zeta}{\zeta^2}\ 2\left(\eta-P_\alpha\right)
\log\left(\eta-P_\alpha\right).
\end{equation}
The K\"ahler potential $K$ is the Legendre transform of $F$:
\begin{equation}
K(a,\bar{a},t,\bar{t})=F-b\left(t+\bar{t}\right), \frac{\partial F}{\partial b}=t+\bar{t}.
\end{equation}
It is a well known fact that the metric corresponding to Eq.~(\ref{F1}) is the multi-Taub-NUT
metric with $k$ centers~\cite{IR,HKLR}. This is in agreement with string theory 
predictions~\cite{us}.

\section{Moduli space of centered $n=2$ monopole}\label{k2}
\subsection{Twistor space $Z_2^0$ of centered $n=2$ monopole}
For $n=2$ the moduli space $M_2$ is 8-dimensional and admits a triholomorphic $U(1)$ action.
We define the centered moduli space $M_2^0$ to be the hyperk\"ahler quotient of $M_2$ with
respect to this $U(1)$ (at zero level). The $U(1)$ action on $M_2$ lifts to a ${\bf C}^*$ action
on $Z_2$. It acts by $\rho_j\ra \lambda\rho_j$, $ \xi_j\ra \lambda^{-1}\xi_j.$ The corresponding
moment map is $\eta_1$, as can be easily seen from the expression for $\omega$. Thus $Z_2^0$,
the twistor space of $M_2^0,$ is the ${\bf C}^*$ quotient of the subvariety $\eta_1=0=\tilde\eta_1$ in $Z_2$.
We first investigate one coordinate patch of $Z_2^0$.
Let us denote $\psi_1=\rho_0\xi_0,\psi_2=\rho_1\xi_1,\psi_3=\frac{1}{2}(\rho_0\xi_1+\rho_1\xi_0),
\psi_4=\frac{1}{2}(\rho_0\xi_1-\rho_1\xi_0)$.
The variables $\psi_i$ are invariant with respect to ${\bf C}^*$ action and satisfy 
\begin{eqnarray}\label{constr2}
&&\psi_1\psi_2=\psi_3^2-\psi_4^2, \\ \nn
&&\psi_1-\eta_2\psi_2+2\sqrt{-\eta_2}\psi_3=\prod_\alpha (\sqrt{-\eta_2}-P_\alpha(\zeta)),\\ \nn
&&\psi_1-\eta_2\psi_2-2\sqrt{-\eta_2}\psi_3=\prod_\alpha (-\sqrt{-\eta_2}-P_\alpha(\zeta)).
\end{eqnarray}
These equations define a three-dimensional subvariety $U^0$ in ${\bf C}^6$ with coordinates 
$(\zeta,\eta_2,\psi_1,\ldots,\psi_4)$. 
Geometric invariant theory tells us that $Z_2^0$ can be obtained by gluing together two copies
of $U^0$ over $\zeta\neq 0,\infty$. The transition functions can be computed from Eq.~(\ref{trk}):
\begin{eqnarray}\label{tr2}
\tilde\zeta&=&\zeta^{-1},\\ \nn
\tilde\eta_2&=&\zeta^{-4}\eta_2,\\ \nn
\tilde\psi_1&=&\frac{\zeta^{-2k}}{2}\left(\psi_1-\eta_2\psi_2+\cos\gamma (\psi_1+\eta_2\psi_2)-
2\psi_4\sqrt{\eta_2}\sin\gamma\right),\\ \nn
\tilde\psi_2&=&\frac{\zeta^{4-2k}}{2\eta_2}\left(-(\psi_1-\eta_2\psi_2)+\cos\gamma (\psi_1+\eta_2\psi_2)-
2\psi_4\sqrt{\eta_2}\sin\gamma\right),\\ \nn
\tilde\psi_3&=&\zeta^{2-2k}\psi_3,\\ \nn
\tilde\psi_4&=&\frac{\zeta^{2-2k}}{2}\left((\psi_1+\eta_2\psi_2)\frac{\sin\gamma}{\sqrt{\eta_2}}+
\psi_4\cos\gamma\right),
\end{eqnarray}
where $\gamma=2\mu\sqrt{\eta_2}/\zeta.$

{}From this explicit description of $Z_2^0$ one can see that for any $\zeta$
the fiber of $Z_2^0$ is a resolution of the $D_k$ singularity. Indeed, combining Eqs.~(\ref{constr2})
we see that the fiber of $U_0$ over $\zeta$ is biholomorphic to a hypersurface in ${\bf C}^3$
(with coordinates $(\eta_2,\psi_2,\psi_4)$) given by 
\begin{equation}
\psi_4^2+\eta_2\psi_2^2+\psi_2{\cal Q}(\eta_2)-{\cal R}(\eta_2)^2=0,
\end{equation}
where ${\cal Q}(\eta_2),{\cal R}(\eta_2)$ are polynomials in $\eta_2$ defined by
\begin{eqnarray}
2{\cal Q}(\eta_2)&=&\prod_\alpha (\sqrt{-\eta_2}-P_\alpha(\zeta))+
\prod_\alpha (-\sqrt{-\eta_2}-P_\alpha(\zeta)),\nn \\
4\sqrt{-\eta_2}{\cal R}(\eta_2)&=&\prod_\alpha (\sqrt{-\eta_2}-P_\alpha(\zeta))-
\prod_\alpha (-\sqrt{-\eta_2}-P_\alpha(\zeta)).\nn 
\end{eqnarray}
Furthermore, these formulas imply that if all points ${\vec p}_1,\ldots,{\vec p}_k$
are distinct, the manifold $M_2^0$ is a smooth complex manifold in any of its complex
structures. Since the 2-form $\omega$ is smooth as well, we conclude that $M_2^0$
is a smooth hyperk\"ahler manifold. The smoothness of $M_2^0$ is also in agreement with string theory
predictions. Indeed, as explained in Ref.~\cite{us}, the space $M_2^0$ is the Coulomb branch
of $N=4,D=3$ $SU(2)$ gauge theory with $k$ fundamental hypermultiplets, with
${\vec p}_\alpha$ being hypermultiplet masses. When ${\vec p}_\alpha$ are all distinct, the theory
has no Higgs branch, and therefore the Coulomb branch is smooth everywhere. When 
some masses become equal, the Higgs branch emerges, and the Coulomb branch develops an 
orbifold singularity at the point where it meets the Higgs branch. Thus we expect
that when some of ${\vec p}_\alpha$ coincide, or equivalently, when some of $\ell_\alpha$ are
bigger than $1$, the manifold $M_2^0$ has orbifold singularities.

In Ref.~\cite{us3} the same manifold $Z_2^0$ arose as the twistor space of the moduli
space of a system of ordinary differential equations (so called Nahm equations).
This is of course a consequence of a general correspondence between solutions of
Bogomolny equations and Nahm equations~\cite{Nahm,H2,HM}. Thus Ref.~\cite{us3} provides an
equivalent construction of $D_k$ ALF metrics.   

\subsection{Real holomorphic sections of $Z_2^0$}

The discussion of section 2 implies that a real holomorphic section of the uncentered 
twistor space
$Z_2$ is a triplet $(S,\rho,\xi)$, where $S$ is the spectral curve in ${\bf T}$ given by
$\eta^2+\eta_1\eta+\eta_2=0$, $\rho$ and $\xi$ are holomorphic sections of $L^{\mu}(k)|_S$ 
and $L^{-\mu}(k)|_S$ satisfying the condition (iii) of section 2. Then, as explained 
in section 3, the real holomorphic sections of $Z_2^0$ are obtained by setting $\eta_1=0$
and modding by the ${\bf C}^*$ action $\rho\to\lambda\rho,\xi\to \lambda^{-1}\xi$. 
In this subsection we find the explicit form of the real holomorphic sections of $Z_2^0$.

The curve $\eta^2+\eta_2=0$ is either elliptic or a union of two ${\bf CP}^1$'s. The former
case is generic, while the latter occurs at a submanifold of the moduli space.
Intuitively the latter
case corresponds to the situation when the two nonabelian monopoles are on top of
each other. It suffices to consider the elliptic case.

By an $SO(3)$ rotation 
\begin{equation}
\zeta=\frac{a\tilde\zeta+b}{-\overline{b}\tilde\zeta+\overline{a}},\qquad
\eta=\frac{\tilde\eta}{(-\overline{b}\tilde\zeta+\overline{a})^2},\qquad |a|^2+|b|^2=1,
\end{equation}
we can always bring the elliptic curve $\eta^2=-\eta_2(\zeta)$ to the form 
\begin{equation}\label{st}
\tilde\eta^2=4k_1^2\left(\tilde\zeta^3-3k_2\tilde\zeta^2-\tilde\zeta\right), k_1>0,k_2\in{\bf R}.
\end{equation}
It follows that the discriminant $\Delta>0$, and therefore the lattice defined by the curve $S$
is rectangular. We denote this lattice $2\Omega$ and its real and imaginary periods by 
$2\omega$ and $2\omega',$ respectively.
 
We parametrized $S$ by five real parameters: the Euler angles of the $SO(3)$ rotation
and a pair of real numbers $k_1$ and $k_2$. We will see in a moment that the condition (iii)
imposes one real constraint on them, so we will obtain a four-parameter family of real 
sections, as required.

To write explicitly a section of $L^{\mu}(k)|_S$, we will use the
standard ``flat'' parameter on the elliptic curve $u$ defined modulo $2\Omega$, in terms of
which $\tilde\eta=k_1{\cal P}'(u),\zeta={\cal P}(u)+k_2.$ Here ${\cal P}(u)$ is the
Weierstrass elliptic function. In terms of $u$ the real structure acts by
$u\ra -\overline{u}+\omega+\omega'.$

A section of $L^{\mu}(k)|_S$ can be thought of as a pair of functions on $S$
$f_1,f_2$ such that $f_1$ is holomorphic everywhere except $\zeta=\infty,$ $f_2$ is
holomorphic everywhere except $\zeta=0,$ and for $\zeta\neq 0,\infty$
$f_2(\zeta)=\zeta^{-k}\exp(-\mu\eta/\zeta)f_1(\zeta).$ The point $\zeta=\infty$ corresponds
to two points $u_\infty,-u_\infty$ on $S$ defined by ${\cal P}(u_\infty)+k_2=
\overline{a}/\overline{b}.$ 
Furthermore, condition (iii) implies that the divisor of $f_1$ is $Q_+$. Let us recall that
$Q_+\bigcup Q_-=\bigcup_\alpha Q_\alpha,$ where $Q_\alpha=S\bigcap P_\alpha, \alpha=1,\ldots,k.$ 
Thus $Q_\alpha$ consists of solutions of a system of two equations $\eta=P_\alpha(\zeta),
\eta^2=-\eta_2(\zeta).$ Obviously, this defines four points on the elliptic curve $S.$ Because
of real structure, these four points split into two pairs whose members are interchanged by
$\tau$. $Q_+$ includes one point from each pair (for all $\alpha$), $Q_-$ includes the 
rest.\footnote{There is a $4^m$-fold ambiguity involved in the splitting 
$Q=Q_+\bigcup Q_-$. It can be fixed, in principle, by requiring that the normal bundle of
every section in the family be ${\cal O}(1)\oplus {\cal O}(1)$.}
Let us denote the ``flat'' coordinates of points in $Q_+$ by 
$u_\alpha,u'_\alpha,\alpha=1,\ldots,k,$ and those in $Q_-$ by 
$v_\alpha,v'_\alpha,\alpha=1,\ldots,k$. By definition, $v_\alpha=-\overline{u}_\alpha
+\omega+\omega'({\rm mod}\ 2\Omega),
v'_\alpha=-\overline{u}'_\alpha+\omega+\omega'({\rm mod}\ 2\Omega).$ We fix the 
${\rm mod}\ 2\Omega$ ambiguity by requiring that $u_\alpha,u'_\alpha,v_\alpha,v'_\alpha$
be in the fundamental rectangle of $2\Omega$.
In this notation a section of $L^{\mu}(k)|_S$ is given by
\begin{equation}
f_1\sim\exp\left(-\mu k_1(\zeta_W(u+u_\infty)+\zeta_W(u-u_\infty))+Cu\right)
\prod_\alpha \frac{\sigma(u-u_\alpha)\sigma(u-u'_\alpha)}{\sigma(u-u_\infty)\sigma(u+u_\infty)}.
\end{equation}
Here $\zeta_W(u)$ and $\sigma(u)$ are Weierstrass quasielliptic functions (we denote Weierstrass
$\zeta$-function by $\zeta_W(u)$ to avoid confusion with the affine coordinate $\zeta$ on the
${\bf P}^1$ of complex structures), and $C$ is a constant.
Similarly, a section of $L^{-\mu}(k)|_S$ with the divisor $Q_-$ is represented by a pair of
functions $g_1,g_2$ related by $g_2(\zeta)=\zeta^{-k}\exp(\mu\eta/\zeta)g_1(\zeta).$
Explicitly $g_1$ is given by
\begin{equation}
g_1\sim\exp\left(\mu k_1(\zeta_W(u+u_\infty)+\zeta_W(u-u_\infty))+Du\right)
\prod_\alpha \frac{\sigma(u-v_\alpha)\sigma(u-v'_\alpha)}{\sigma(u-u_\infty)\sigma(u+u_\infty)},
\end{equation}
where $D$ is another constant.
In general $f_1$ and $g_1$ are quasiperiodic with periods $2\omega$ and $2\omega'$. The condition
(iii) is equivalent to asking that $f_1$ and $g_1$ be doubly periodic.
One can see that the latter can be achieved by adjusting $C$ and $D$ if and only if
\begin{eqnarray}\label{latt}
&&2\mu k_1+\sum_\alpha(u_\alpha+u'_\alpha)\in 2\Omega,\nn\\
&&2\mu k_1-\sum_\alpha(v_\alpha+v'_\alpha)\in 2\Omega.
\end{eqnarray}
Recalling that $k_1$ is real and positive, we conclude that there exist integers
$m,m',p,p'$ and a real number $x\in \left(0,2\omega\right]$ such that
$\sum_\alpha(u_\alpha+u'_\alpha)=-x+2m\omega+2m'\omega',
\sum_\alpha(v_\alpha+v'_\alpha)=x+2p\omega+2p'\omega'$.
Then Eqs.~(\ref{latt}) together with the condition (iv) imply 
\begin{equation}\label{cond}
2\mu k_1=x.
\end{equation} 
Then for $f_1$ and $g_1$ to be doubly periodic one has to set
$$C=2m\zeta_W(\omega)+2m'\zeta_W(\omega'), D=2p\zeta_W(\omega)+2p'\zeta_W(\omega')$$
Let us notice for future use that 
\begin{eqnarray}\label{fperiod}
&&\log f_1(u+\omega)-\log f_1(u)=-2\pi im',\nn\\
&&\log f_1(u+\omega')-\log f_1(u)=2\pi im,\nn\\
&&\log g_1(u+\omega)-\log g_1(u)=-2\pi ip',\nn\\
&&\log g_1(u+\omega')-\log g_1(u)=2\pi ip.
\end{eqnarray}

Eq.~(\ref{cond}) is a transcedental equation on $k_1,k_2,$ and the $SO(3)$ rotation
required to bring $S$ to the standard form Eq.~(\ref{st}). It reduces the number of real
parameters in the equation of the curve from 5 to 4. Thus we have a four-parameter
family of real sections of $Z_2^0$.

\subsection{The K\"ahler potential of the centered $n=2$ moduli space}

Having found a four-parameter family of real holomorphic sections of $Z^0_2$ we now would like to compute
the corresponding hyperk\"ahler metric. Since $Z^0_2$ has an intermediate holomorphic projection on
${\cal O}(4),$ we can use the method of Ref.~\cite{IR} to write down
the Legendre transform of the K\"ahler potential.
The existence of the projection is equivalent to saying that $\eta_2$ is a holomorphic coordinate on 
$Z^0_2$.
The holomorphic 2-form $\omega$ in the patch $\zeta\neq\infty$ can be written as
$$\omega=d\eta_2\wedge d\sum_{branches}{\frac{1}{\eta}\log \frac{f_1}{g_1}}\equiv d\eta_2\wedge
d\chi.$$
Here $f_1$ and $\eta=\sqrt{-\eta_2}$ are regarded as double-valued functions of
$\zeta\in {\bf P}^1\backslash\{\zeta\neq\infty\}$, 
and the sum is over the two branches of the cover $S\ra {\bf P}^1$. Similarly,
in the patch $\zeta\neq 0$ we can write
$$\omega'=d\eta_2'\wedge d\chi'.$$
On the overlap we have the relations
\begin{equation}\label{transchi}
\omega'=\zeta^{-2}\omega,\quad\eta_2'=\zeta^{-4}\eta_2,
\quad\chi'=\zeta^2\chi-4\mu\zeta.
\end{equation}

Following Ref.~\cite{IR}, we would like to find a (multi-valued) function $\hat{f}(\eta,\zeta)$
and a contour $C$ on the double cover $S\ra {\bf P}^1$ such that 
$$\oint_C\frac{d\zeta}{\zeta^j}\hat{f}(\eta,\zeta)=\oint_0\frac{d\zeta}{\zeta^{j-2}}
\chi+\oint_\infty\frac{d\zeta}{\zeta^j}\chi'$$
for any integer $j$. Here the contours of integration
on the RHS are small positively oriented loops around $\zeta=0$ and 
$\zeta=\infty$. To find $\hat{f}$ we substitute the explicit expressions for $\chi$ and $\chi'$
and rewrite the integral on the RHS as an integral in the $u$-plane. Then the RHS
becomes
\begin{equation}\label{step1}
\oint \frac{du}{k_1} \zeta(u)^{-j+2} \log \frac{f_1(u)}{g_1(u)}+4\mu\oint_0\frac{d\zeta}{\zeta^{j-1}},
\end{equation}
where the contour in the first integral consists of four small positively oriented
loops around four preimages of the points $\zeta=0$ and $\zeta=\infty$ in the 
fundamental rectangle of the lattice $2\Omega.$ We denote these
points $u_0,u'_0=2(\omega+\omega')-u_0,u_\infty,u'_\infty=2(\omega+\omega')-u_\infty.$
Besides these four points the only other branch points of $\log f_1(u)/g_1(u)$ in the
fundamental rectangle are $u_\alpha,u'_\alpha,v_\alpha,v'_\alpha, \alpha=1,\ldots,k.$
As for $\zeta(u)$, it is elliptic. Then we can rewrite Eq.~(\ref{step1}) as
\begin{equation}\label{integration1}
\oint_{bdry} \frac{du}{k_1} \zeta(u)^{-j+2} \log \frac{f_1(u)}{g_1(u)}+
\sum_\alpha\oint_{A_\alpha+A'_\alpha}
\frac{du}{k_1}\ \zeta(u)^{-j+2} \log \frac{f_1(u)}{g_1(u)}+
4\mu\oint_0\frac{d\zeta}{\zeta^{j-1}},
\end{equation}
where the contour in the first integral runs along the boundary of the fundamental
rectangle, while $A_\alpha$ and $A'_\alpha$ enclose the pairs of points $u_\alpha,v_\alpha$
and $u'_\alpha,v'_\alpha$, respectively (see Figure 2). 
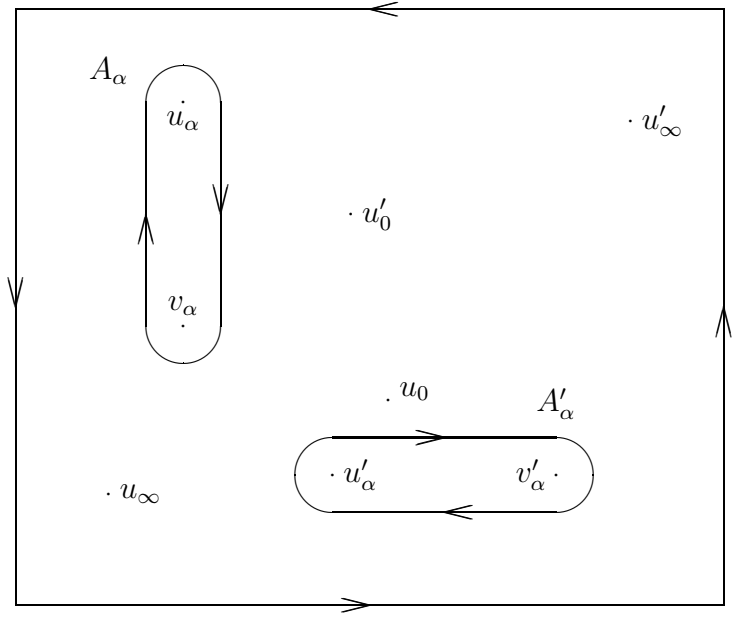
\begin{figure}[t]\label{fig2}
\setlength{\unitlength}{0.6em}
\begin{center}

\begin{picture}(40,34)
\put(10,22){\oval(4,16)} 
\put(8,22){\line(-1,-4){0.4}} \put(8,22){\line(1,-4){0.4}}
\put(12,22){\line(-1,4){0.4}} \put(12,22){\line(1,4){0.4}}
\put(10,28){\circle*{0.2}} \put(10,16){\circle*{0.2}}
\put(10,28){\makebox(0,-0.5)[t]{$u_\alpha$}} 
\put(10,16.6){\makebox(0,0)[b]{$v_\alpha$}}
\put(6,29){\makebox(0,0)[b]{$A_\alpha$}}

\put(24,8){\oval(16,4)}
\put(24,10){\line(-4,1){1.5}} \put(24,10){\line(-4,-1){1.5}}
\put(24,6){\line(4,1){1.5}} \put(24,6){\line(4,-1){1.5}}
\put(18,8){\circle*{0.2}} \put(30,8){\circle*{0.2}} 
\put(18.6,8){\makebox(0,0)[l]{$u'_\alpha$}}
\put(30,8){\makebox(-0.5,0)[r]{$v'_\alpha$}}
\put(6,7){\circle*{0.2}} \put(6.6,7){\makebox(0,0)[l]{$u_\infty$}}
\put(34,27){\circle*{0.2}} \put(34.6,27){\makebox(0,0)[l]{$u'_\infty$}}
\put(21,12){\circle*{0.2}} \put(21.6,12){\makebox(0,0)[lb]{$u_0$}}
\put(19,22){\circle*{0.2}} \put(19.6,22){\makebox(0,0)[l]{$u'_0$}}
\put(30,11){\makebox(0,0)[b]{$A'_\alpha$}}

\put(1,1){\line(1,0){38}} \put(39,1){\line(0,1){32}}
\put(39,33){\line(-1,0){38}} \put(1,33){\line(0,-1){32}}
\put(20,1){\line(-4,-1){1.5}} \put(20,1){\line(-4,1){1.5}}
\put(39,17){\line(-1,-4){0.4}} \put(39,17){\line(1,-4){0.4}}
\put(20,33){\line(4,1){1.5}} \put(20,33){\line(4,-1){1.5}}
\put(1,17){\line(-1,4){0.4}} \put(1,17){\line(1,4){0.4}}


\end{picture}\end{center}
\caption{Integration contours in Eq.~(\ref{integration1}).
Only one of the contours $A_\alpha$ and one of the contours $A'_\alpha$
are shown.}
\end{figure}
Using Eqs.~(\ref{fperiod}) the integral over the boundary can be simplified to
$$-2\pi i\oint_{(m-p,m'-p')} \frac{du}{k_1}\ \zeta(u)^{-j+2},$$
where the contour $(m-p,m'-p')$ winds $m-p$ times around the real cycle and $m'-p'$ times 
around the imaginary cycle. Recalling the explicit form of $f_1(u)$ and $g_1(u)$, we can 
rewrite the integral over $A_\alpha+A'_\alpha$ as 
\begin{equation}\label{interm}
\oint_{B_\alpha+B'_\alpha}\frac{du}{k_1} \zeta(u)^{-j+2} \log \sigma(u-u_\alpha)\sigma(u-u'_\alpha)
\sigma(u-v_\alpha)\sigma(u-v'_\alpha),
\end{equation}
where the contours $B_\alpha$ and $B'_\alpha$ are figure-eight-shaped contours shown in
Figure 3. 
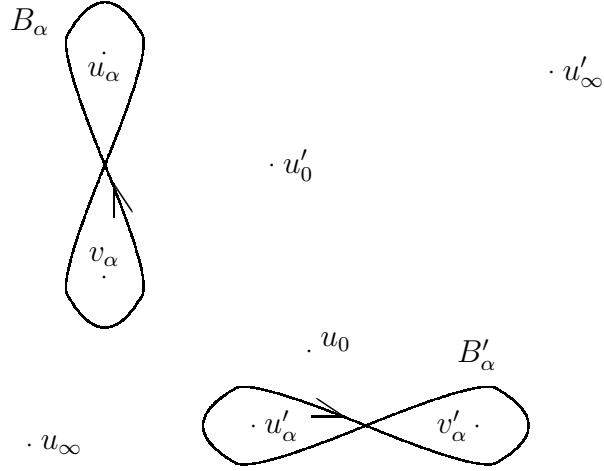
\begin{figure}[t]\label{fig3}
\setlength{\unitlength}{0.6em}
\begin{center}

\begin{picture}(40,34)
\qbezier(8,29)(10,32.464)(12,29)
\qbezier(8,29)(7.4896,28.116)(10,22)
\qbezier(12,29)(12.5104,28.116)(10,22)
\qbezier(12,15)(12.5104,15.884)(10,22)
\qbezier(8,15)(7.4896,15.884)(10,22)
\qbezier(8,15)(10,11.536)(12,15)
\put(10.5,21){\line(2,-3){1.0}} \put(10.5,21){\line(0,-1){1.8}}
\put(10,28){\circle*{0.2}} \put(10,16){\circle*{0.2}}
\put(10,28){\makebox(0,-0.5)[t]{$u_\alpha$}} 
\put(10,16.6){\makebox(0,0)[b]{$v_\alpha$}}
\put(6,29){\makebox(0,0)[b]{$B_\alpha$}}

\qbezier(17,10)(13.536,8)(17,6)
\qbezier(17,10)(17.884,10.5104)(24,8)
\qbezier(17,6)(17.884,5.4896)(24,8)
\qbezier(31,10)(30.116,10.5104)(24,8)
\qbezier(31,6)(30.116,5.4896)(24,8)
\qbezier(31,10)(34.464,8)(31,6)
\put(22.9,8.5){\line(-3,2){1.5}} \put(22.9,8.5){\line(-1,0){1.8}}
\put(18,8){\circle*{0.2}} \put(30,8){\circle*{0.2}} 
\put(18.6,8){\makebox(0,0)[l]{$u'_\alpha$}}
\put(30,8){\makebox(-0.5,0)[r]{$v'_\alpha$}}
\put(30,11){\makebox(0,0)[b]{$B'_\alpha$}}

\put(6,7){\circle*{0.2}} \put(6.6,7){\makebox(0,0)[l]{$u_\infty$}}
\put(34,27){\circle*{0.2}} \put(34.6,27){\makebox(0,0)[l]{$u'_\infty$}}
\put(21,12){\circle*{0.2}} \put(21.6,12){\makebox(0,0)[lb]{$u_0$}}
\put(19,22){\circle*{0.2}} \put(19.6,22){\makebox(0,0)[l]{$u'_0$}}
\end{picture}\end{center}
\caption{The contours $B_\alpha$ and $B'_\alpha$.}
\end{figure}
On the other hand, it can be easily seen that
$$\eta(u)-P_\alpha(\zeta(u))\sim e^{u(C+D)}\ 
\frac{\sigma(u-u_\alpha)\sigma(u-u'_\alpha)\sigma(u-v_\alpha)\sigma(u-v'_\alpha)}{
\sigma(u-u_\infty)^2\sigma(u+u_\infty)^2}.$$
Since neither $u_\infty$ nor $u'_\infty$ are enclosed by the contour $B_\alpha+B'_\alpha$, the
integral Eq.~(\ref{interm}) is equal to
$$\oint_{B_\alpha+B'_\alpha}\frac{du}{k_1} \zeta(u)^{-j+2} \log (\eta(u)-P_\alpha(\zeta(u))) .$$
Collecting all of this together we get
\begin{eqnarray}\label{hatf}
\oint_C\frac{d\zeta}{\zeta^j}\hat{f}(\eta,\zeta)\!\!&=&\!\!-2\pi i\oint_{(m-p,m'-p')}\frac{d\zeta}{\eta}
\zeta^{-j+2}\\ \nn
& &\mbox{}\!\!+\sum_\alpha\oint_{C_\alpha+C'_\alpha}\frac{d\zeta}{\eta}\zeta^{-j+2} \log(\eta-P_\alpha(\zeta))
+4\mu\oint_0\frac{d\zeta}{\zeta^{j-1}}.
\end{eqnarray}
Here all the functions are regarded as functions on the double cover of the $\zeta$-plane,
and the contours $C_\alpha,C'_\alpha$ are the images of $B_\alpha,B'_\alpha$ under the map
$u\mapsto\zeta.$ We now define a function $G(\eta,\zeta)$ by $\partial G/\partial \eta=-2\eta\zeta^{-2}\hat{f}$.
According to Ref.~\cite{IR} the Legendre transform of the K\"ahler potential is given by
$$F=\frac{1}{2\pi i}\oint \frac{d\zeta}{\zeta^2} G(\eta,\zeta).$$ Hence we can read off $F$:
\begin{eqnarray}\label{F}
F&=&-\frac{1}{2\pi i}\oint_0 d\zeta \frac{4\mu\eta^2}{\zeta^3}+\oint_{(m-p,m'-p')}\ d\zeta
\frac{2\eta}{\zeta^2} \\ \nn
 & & \mbox{}-\sum_\alpha\frac{1}{2\pi i}\oint_{C_\alpha+C'_\alpha}\frac{d\zeta}{\zeta^2}\,
2(\eta-P_\alpha(\zeta))\log (\eta-P_\alpha(\zeta)).
\end{eqnarray}
$F$ may be regarded as a function of the coefficients of $\eta_2(\zeta)=z+v\zeta+w\zeta^2-
\overline{v}\zeta^3+\overline{z}\zeta^4.$ Since $w$ is real, $F$ depends on 5 real parameters.
These parameters are subject to one transcedental constraint expressed by Eq.~(\ref{cond}).
(This constraint implies $\partial F/\partial w=0.$) Thus we may think of $w$ as an 
implicit function of $z$ and $v$. The K\"ahler potential $K(z,\overline{z},u,\overline{u})$
is the Legendre transform of $F$:
$$K(z,\overline{z},u,\overline{u})=F(z,\overline{z},v,\overline{v},w)-uv-
\overline{u}\overline{v},\ \
\frac{\partial F}{\partial v}=u,\frac{\partial F}{\partial \overline{v}}=\overline{u}.$$
Eq.~(\ref{F}) agrees with a conjecture by Chalmers~\cite{Ch}.

We already saw in section 5 that $M_2^0$ is a resolution of $D_k$ singularity. Now we can
check that it is ALF. To this end we take the limit $k_1\to +\infty$. Eq.~(\ref{cond})
implies that in this limit $\omega\to\infty$, while $\omega'$ stays finite. Thus the
curve $S$ degenerates: $\eta_2(\zeta)\to-(P(\zeta))^2$, where $P(\zeta)$ is a real section
of ${\bf T}$. It is easy to see that in this limit $F$ reduces to the Taub-NUT form (see
section 4):
\begin{equation}
F\sim \frac{1}{2\pi i}\oint_0 d\zeta \left(-\frac{4\mu P(\zeta)^2}{\zeta^3}+
K \frac{P(\zeta)\log P(\zeta)}{\zeta^2}\right),
\end{equation}
where $K$ is an integer depending on the limiting behavior of $u_\alpha,u'_\alpha,v_\alpha,v'_\alpha$.
Therefore asymptotically the metric on $M_2^0$ has the Taub-NUT form. With some more work it
should be possible to compute the integer $K$ as well. 

Note also that if we set $\mu=0$, then the metric becomes ALE. Kronheimer proved~\cite{Torelli}
that the $D_k$ ALE metric is essentially unique. Thus we have obtained the Legendre transform of the
K\"ahler potential for the $D_k$ metrics of Ref.~\cite{ALE}. It would be interesting to obtain
a similar representation for the $E_k$ ALE metrics.

\end{document}